\documentclass[preprint,prx,longbibliography]{revtex4-1}
\usepackage{subcaption}
\usepackage{graphicx}
\usepackage{comment}
\usepackage{amsmath,amssymb,amsthm} 
\usepackage{scalerel,stackengine}
\usepackage{bm}
\usepackage{verbatim}
\usepackage{float}
\begin{document}
\let\WriteBookmarks\relax
\def\floatpagepagefraction{1}
\def\textpagefraction{.001}

\title { Origin of homochirality in peptides: The first milestone at the origin of life }                      
\newcommand{\orcidauthorA}{0000-0002-0819-698X} 
\author{S\o ren Toxvaerd}
\affiliation{DNRF centre  ``Glass and Time'',  Department of Science and Environment, Roskilde University, Postbox 260, DK-4000, Denmark}
\email{st@ruc.dk; doi.org/10.1016/j.biosystems.2025.105479 }

\begin{abstract} 
 Living organisms have some common structures,
chemical reactions and molecular structures.
 The organisms consist of cells with cell division, they have homochirality of protein and carbohydrate units, metabolism, and genetics, and they are mortal.
The molecular structures and chemical reactions underlying these features are common to all, from the simplest bacteria to human beings.
The origin of life is evolutionary with the emergence of a network of spontaneous biochemical reactions, and the evolution has taken place over a very long time.
 The evolution contains, however, some ``landmarks'' and bottlenecks, which in a revolutionary manner directed the evolution,
and the article establishes an order of some of these events.
Recent articles show  that peptides in living organisms are long-time unstable with loss of their secondary homochiral conformations and with
D-amino acids.
Based on these observations and an extensive scientific literature on Abiogenesis, we argue 
that the first milestone in the prebiotic evolution is at the emergence of homochiral peptides in an aqueous solution with a high concentration of amino acids and a lower 
water activity than in the cytosol in living organisms. The homochiral peptides in cytosol are unstable, and the long-time aging of peptides in the cytosol
causes mortality of living organisms. The metabolism and genetics are established in an environment with homochiral peptides in the Earth's crust for 
$\approx$ 4 Gyr ago at a lower water activity than in the cytosol in living organisms.
	Finally, the cells with cell division are established in the Hot Springs environment at the interface between the crust and the
Hadean Ocean.
\end{abstract}


\maketitle

\section{Introduction}
There have been living organisms on Earth for at least  3.5  billion  years \cite{Djokic2017,Schopf2018,Cavalazzi2021,Dodd2017} and all 
living organisms have some common features. 
The molecular structures and chemical reactions underlying these common features are necessary for all living organisms, from the simplest bacteria to human beings.
The Earth is $\approx$ 4.5 billion years old, and if life has originated here these  molecular structures and chemical reactions have been established in a
prebiotic environment in the time spent before the emergence of the first primitive living organisms. The article assumes that the preconditions for establishing
living organisms have been established in  prebiotic environments  before the first appearance of a  bacteria
$\approx$ 3.5 billion years ago, and that these common  molecular structures and chemical reactions in living organisms 
have  been established by exergonic (spontaneous) reactions
over many millions of years. 

Some common features of living organisms are:\\
1. Living organisms consist of cells with content, which from a physicochemical point of view are soft condensed
matter in a $diluted$ aqueous suspension. \\ 
2. The cells have cell division.\\
3. Homochirality of protein and carbohydrate units, and stereospecific molecules in genetics.\\
4. Metabolism. \\
5. The genetics. \\
6. Mortality. All living organisms suffer from aging  and are mortal.\\ 
.... and there are many other general and important features, ignored or overlooked here.

The establishment of common features of living organisms raises  a philosophical question:
 Is the origin of life a consecutive series of revolutionary events where the next step in evolution is only possible due to previous
 events? Or it could be that Abiogenesis is the result of a gradual 
 process with increasing complexity of molecules, chemical
reactions, and confinements in cells? 
 The causality dilemma is formulated as a philosophical question: ``The chicken or the egg  dilemma, which came first: the chicken or the egg?"
 The answer to the question is that we expect causality in the life cycle at a time when it is established, and the problem
 is to establish a possible causality at the beginning of life.

There exist two forms of simple living cells: bacteria and archaea, which mainly differ
by the constitutions of their membranes.
Their metabolism, genetics, and enzyme systems are qualitatively the same, and they are both associated with homochiral
 peptide and carbohydrate units and with an interior of the cells, which is a diluted aqueous suspension.
These facts seem to indicate, that prebiotic biosynthesis, metabolism, and genetics were established 
before life with cells was established. Furthermore, a cell with the biomolecules, necessary for a living system is a structure
that either is established or not, a decisive event and a milestone in the Abiogenesis. This fact points to the
answer to the question at least partly is, that Abiogenesis is a consecutive sequence of reactions with milestones
at decisive events at the origin of life. However, evolution is most likely
 a combination of these decisive events and a gradual process with the increasing complexity of molecules and structures. This
enabled the next step in a revolutionary evolution.

This article advocates that the evolution is partly revolutionary with 
a sequence of milestones at a series of  decisive events, and the  article tries to establish an order in some of these events. 

\section{ Milestones in the Abiogenesis}

 A cell is a ``space capsule'' that protects its content from the exterior, and a living cell protects its life with the interior biomaterials from
 the exterior aqueous environment which differs physicochemically from the environment in the cell \cite{Toxvaerd2023}.
 Hence, establishing the cell and cell division is likely with the last milestone
 at the origin of life. The order in the sequence of some of the decisive events might be:

A. Emergence of homochirality in the units in proteins and mortality of life.

B1. Emergence of a metabolism.

B2. Emergence of the genetics.

C) Emergence of cells with spontaneous cell division.

The rationale for this order in the Abiogenesis  with milestones at decisive events is given below.

\subsection{A. Emergence of homochirality and razemization of the units in proteins} 

There are an overwhelming number of articles that deal with the origin of homochirality in biomolecules and for good reason.
The basic spontaneous, non-equilibrium bio-reactions in a cell are governed by stereospecific enzymes, and the genetic codes in DNA and RNA
are obtained by units with a stereospecific structure. Without stereospecificity, no life is like ours, and it is urgent to maintain the
stereospecificity in a cell.
For recent reviews about the origin of homochirality see  \cite{Blackmond2019,Bunse2021,Cowan2022,Gagnon2023}.

However, the articles about homochirality in peptides overlook a crucial property of amino acids and peptides.

\textit{Amino acids and peptides
are unstable concerning chirality. 
 An aqueous solution of amino acids is racemic due to an active isomerization kinetics } \cite{Bada1972,Bada1985}.  

The racemization of amino acids ranges from a thousand years at low temperatures
to days at 100 $^\circ C$, and the racemization is enhanced in acidic and basic solutions.
Some amino acid units in peptides in 
 the secondary structure also loses their L-structure  \cite{Fujii2010}. 
So, although it is interesting whether amino acids and other biomolecules are brought to  Earth by meteors,
and whether the amino acids in the meteors were in a homochiral configuration or not,
or whether homochirality was obtained by an enantioselective synthesis, or by another  induced  asymmetry ordering, the problem is:

\textit{How to get homochirality in peptides in an aqueous suspension and to maintain the homochiral peptides
	in a relevant biochemical aqueous environment for millions of years, sufficient for the
	establishment of more complex homochiral structures?}

The peptides in living organisms are almost homochiral concerning their units of  L-amino acids,
but some peptides contain a minor amount of D-amino acids \cite{Genchi2017}. A solution of amino acids will racemize. However,
a synthesis of homochiral peptides from an aqueous racemic  solution of amino acids can be obtained
by consecutive reactions, first suggested by C. F. Frank  \cite{Frank1953}. 
In 1953  he published a theoretical model for spontaneous (i.e. exergonic: $ \Delta G_r(aq)< 0$) $asymmetric$ synthesis, and the
 polymerization of peptides is an example of such a synthesis provided it is  exergonic with a negative
 reaction Gibbs energy,  $ \Delta G_r(aq)< 0$, for the consecutive reactions.

 Since then, many articles have been published on asymmetric polymerization and chiral amplification.  \cite{Brandenburg2005,Ribo2017,Blanco2017,Buhse2022,Pineros2022}.
 As mentioned, it is important to include the racemization of the monomers in the chemical network of
 consecutive reactions, as is the case in several theoretical reaction models  \cite{Ribo2017,Blanco2017,Buhse2022,Pineros2022}.
The coupled reactions are solved numerically for different values of the reaction constants for the
individual chemical reactions in the network.

\begin{figure}
\centering
\label{Figure 1.}
\includegraphics[width=0.80\linewidth]{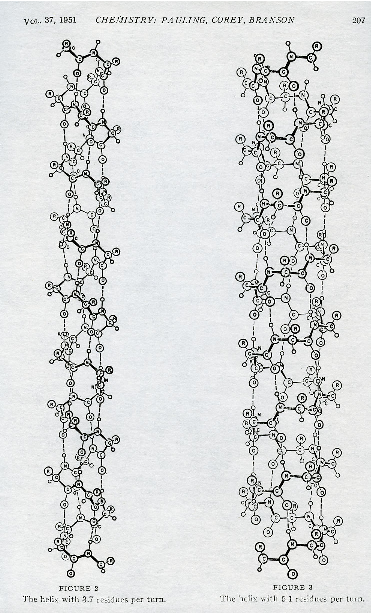}
\vspace{6pt}
\caption{The secondary $\alpha$-helix  structure of a homochiral peptide chain \cite{Pauling1951}. The secondary structure is stabilized by
	(weak) hydrogen bonds. Each amide group is hydrogen-bound to the third (or fifth) amide group beyond
	it along the helix. The  $\alpha$-helix structure to the left of the figure is found in nature.}
\end{figure}

The structure of a bioactive homochiral peptide is given by the
 secondary structure. It was first determined by Pauling \cite{Pauling1951}, and Pauling's figure of the $\alpha-$helix
 illustrates the condition for stability of a homochiral peptide in the aqueous suspension (cytosol)
 in a living cell.
The $\alpha-$helix structure (left) in Figure 1 with homochiral amide units of amino acids was deduced using
spectroscopic data for bond lengths and angles for the 
planar peptide units, and with an $\it{intramolecular}$ stabilizing hydrogen bond
between a hydrogen atom in the substituted -N-H group and an oxygen atom in a
-C=O group in the helix. Only for homochiral amino acids does one get the helix conformation as the overall orientation is changed by a constant
angle at each peptide bond. The $\alpha-$helix structure of an all L-peptide is destroyed if an
L-conformation changes to a D-conformation in  a unit in the peptide, and the peptide loses the secondary structure \cite{Fujii2010}.
Pauling also predicted the second secondary structure of
peptides, the $\beta-$sheet conformation \cite{Pauling1951a}, and this structure is also stabilized by similar hydrogen bonds between parallel sheets.

 The thermodynamic condition for obtaining these secondary protein structures is given in \cite{Toxvaerd2023}. It is due to
``chiral discrimination'' where the gain of enthalpy,  $ \Delta H_r(aq)< 0$,  exceeds the loss
, $ -T\Delta S_r(aq)> 0$, of reaction Gibbs free energy, by the entropy loss due to the ordered protein structure. The enthalpy gain is obtained
by intramolecular hydrogen `` bonds'' between the  -N-H groups and the amide bonds (see Figure 1).
However, the energy of a -N-H$\cdot\cdot\cdot$O=C- hydrogen bond ($\approx$ 10 kJ/mol) is only
of the order of one-half of the energy of a
hydrogen bond between oxygen and hydrogen in pure water \cite{Wendler2010}, and the stabilizing intramolecular hydrogen bonds compete with
hydrogen bonds between the  -N-H groups and oxygens in the water molecules in the solvent.  The 
 -N-H groups in the helix will have a higher tendency to perform hydrogen bonds
 with water molecules if the protein is surrounded with pure water,   and the $\alpha-$helix structure will be
 destabilized in peptide suspensions in pure water. The protein in a pure water suspension will over time lose its helix structure,  but
the secondary structure will be more stable if the water activity is reduced, e.g. by ions Na$^+$, K$^+$, Ca$^{2+}$,HPO$_4^{2-}$,H$_2$PO$_4^{-}$,...
present in the aqueous cytosol in  a living cell.

\subsubsection{Proteins in the cells}
 Today the proteins are mainly present in the cytoplasm in the cells.
The cytoplasm is a highly ordered ''gel-like'' fluid \cite{Bonucci2023}, where the water is mainly located in the cytosol which
contains about $\approx$ 70 \% of the cytoplasm.
The water activity within the cytosol is equal to the activity
in bulk water  \cite{Tros2017}, and the small concentrations of ions in the cytosol must be characterized as a diluted ionic solution \cite{Toxvaerd2023}.
Most proteins are located within the densely packed part of the cytoplasm with a reduced water concentration
which, however,  does not slow down the mobility of the protein molecules \cite{Huang2022}. The hypothesis is that the homochiral peptides were spontaneously
synthesized by an exergonic reaction in a prebiotic environment with low water activity. The peptides and the proteins are now mainly present in the cytoplasm in the cells. They are unstable and can racemize
due to too-high water activity. 

 There is  experimental evidence that water activity is crucial for the stability of the proteins.
  The structure and dynamics of the protein  as a function of the water activity are reviewed in \cite{Bellissent-Funel2016}.
 Simulations of models for polypeptides in a solvent with different
 water activities  show, that
a decreasing activity of an aqueous solvent 
stabilizes the compact conformation of the peptide and it indicates that homochirality
 is obtained  in an aqueous solution with a high ionic concentration or low water activity\cite{Toxvaerd2017,Toxvaerd2023}.

  The secondary peptide conformation is no longer stable at high water activity. This explains that bacteria have an increased
 lifespan at reduced water activity in the bacteria \cite{Liu2018}.
 The peptides in bacteria
are long-time unstable, especially due to the aspartic acid units \cite{Aki2013,Fujii2018,Liang2019,Onstott2014,Liang2021a,Morvan2023}.
The homochirality in the proteins and its stability in bacteria  is vital
for the survival of living bacteria \cite{Jin2020}. One of the most resistant living organisms is spore-forming bacteria where endospores at low temperatures 
 can survive for perhaps millions of years.
 The physical state of water in bacterial spores is reviewed in \cite{Sunde2009}, spores have reduced content of (mobile) water which ensures the secondary structures
of peptides.
 But at room temperature or above the racemization destroys the L-order of aspartic acid units and
 the endospores dies 
\cite{Liang2019}. 

\subsubsection{Aging of a living cell}
Fujii suggested, based on observations of D-Asp in aged cells, that the presence
of  D-Asp is a measure of aging  \cite{Fujii2018}.
However, aging in living organisms, now 3-4 billion years after the emergence of life,
is a much more complex process and is associated with other decays of vital mechanisms for ensuring their life \cite{Mikula2021,Gladyshev2021}, such as the
ability to perform cell division. But even humans are directly
affected  by the homochiral instability \cite{Abdulbagi2021}, e.g. the proteins in Alzheimer's disease contain
a substantial amount of D-aspartic acid units \cite{Liu2023} and the presence of D-amino acids in a living cell
has been linked to schizophrenia, amyotrophic lateral sclerosis, age-related cataracts, and atherosclerosis \cite{Bastings2019}.
The racemization and aging of proteins are reviewed in \cite{Dyakin2021}.

\textit{ In conclusion}: The emergence of homochiral peptides in aqueous \textit{in vitro} suspensions can be established by spontaneous exergonic
(Frank) polymerization of amino acids in a prebiotic aqueous environment with high ionic concentration and a small water activity compared with the water activity in bacteria.
The homochirality by chiral discrimination is achieved by the secondary  $\alpha-$helix  and $\beta-$sheet  structures that require homochirality \cite{Pauling1951,comment}.
 But the secondary peptide conformations are unstable
 in the aqueous environment in the bacteria, especially  the units of aspartic acid  
which racemize. 
All  living organisms on Earth are experiencing this instability and aging, and thus the milestone for the emergence of
homochirality in peptides is also a milestone for the aging and mortality of living organisms.

\subsection{ 
 B. Emergence of a  metabolism and  a genetics}

A living organism has a metabolism, which ensures the necessary free Gibbs energy for the genetics,  cell division, and many other complex
 non-equilibrium endergonic chemical reactions in a living organism. The living organism has also a genetics that
 ensures a unique consecutive network of reactions with stereospecifique molecules, using enzymes. 

 Molecules with homochiral polymer units of L-amino acids and D-carbohydrates
 are common for metabolism and genetics. Another common feature of
 the stereochemistry for carbohydrates and peptides is the fact that the polymer units, simple D-carbohydrates and L-amino acids, are unstable
 and with spontaneous isomerization to racemic compositions. The isomerization kinetics for
 mono-carbohydrates, e.g., Glyceraldehyde and Ribose, and amino acids, is rapid, especially for the simple carbohydrates which tend to
 racemic compositions within hours or days.
 The kinetics for Glysealdehyde is described in  \cite{Fedoro1969}, Ribose isomerizes rapidly by a Maillard kinetics \cite{Laroque2008}.
 So the homochirality of the carbohydrate polymers in  Glycolysis and  RNA
 can not be obtained from polymerizations of homochiral units but must have been obtained in another way.
 The homochirality of peptides can, as pointed out in the previous section, be obtained by the secondary conformational structure of proteins by chiral discrimination,
 but the carbohydrates in the metabolism of living organisms
 are not characterized by having a unique secondary structure. However, the polymers with carbohydrate units in the genetics, DNA and RNA, are characterized by having
 secondary structures with D-carbohydrate units, but RNA is not long-time stable \cite{Kornienko2024}.
 So we must search for another mechanism for the emergence and preservation of
 homochirality of carbohydrates with  D-carbohydrate units (D-Ribose, D-2-deoxyribose,
and D-Glucose).

 A more detailed physicochemical description of the central chemical reactions
 in the metabolism, the Glycolysis, is given in the
 next subsection.  The Glycolysis consists of a network of consecutive and reversible chemical reactions which, for some central chemical reactions are
 catalyzed by stereo-specific enzymes for D-carbohydrates. However, these enzymes are generated by the genetics so the problem about the order of
 synthesis of prebiotic bio-materials is again the ``chicken or the egg'' dilemma.  
 The evolutions of metabolism and genetics were most likely evolutionary,
 but both establishments deserve a milestone at the times of their appearances,
 and before the next step in an evolutionary and revolutionary  prebiotic evolution could occur.

 \subsubsection{ B1. Emergence of the  metabolism }
 Carbohydrates and derivatives of carbohydrates in biosystems are homochiral with respect to the
  D-configuration  at a sertain carbon-unit. The  D-configuration refers to the
 stereo specific configuration of  ligands at the  carbon atom next to the
terminal  -CH$_2$OH for carbohydrates. The homochirality is crucial
for the coupled biochemical reactions in the Glycolysis which is a common metabolism in living systems.

The Formose reaction synthesizes carbohydrates, 
and it is the spontaneous  condensation of formaldehyde
 into carbohydrates,  which was discovered already in 1861 \cite{Butlerow}.
Formaldehyde can be synthesized from simple molecules in the prebiotic environment \cite{Cleaves2008}, and the first steps
in the Formose reaction are the synthesis of formaldehyde into glycolaldehyde and  glyceraldehyde \cite{Appayee2014}
 The first chiral molecule in
the Formose reaction is glyceraldehyde, and the corresponding key molecule in the Glycolysis is
glyceraldehyde-3-phosphate.  Glyceraldehyde has, however, an active isomerization kinetics in an aqueous solution, and 
 this will, without other chemical constraints, give a racemic composition at relevant biological concentrations \cite{Fedoro1969}.
The \textit{in vivo} synthesis of carbohydrates is with enzymes, but the   \textit{in vitro} Formose synthesis of carbohydrates results in
racemic compositions \cite{Weber2001}. The carbohydrates have no secondary structures with a stabilization of all D- or L- conformation so the
emergence of homochirality in carbohydrates is not ensured by chiral discrimination in a secondary conformation.

 Living organisms synthesize only  D-carbohydrates in the metabolism,  Glycolysis.
 However, Glycolysis contains a very effective enzyme, triose phosphate isomerase.
 The enzyme eliminates the activation energy in the isomerization kinetics of glyceraldehyde-3-phosphate and ensures a racemic composition 
 of D-glyceraldehyde-3-phosphate and L-glyceraldehyde-3-phosphate \cite{Alberty}. This
simple enzyme   is believed to have been present in LUCA (Last Universal Common Ancestor)
 and could be among the very
first enzymes in the evolution  \cite{Sobolevsky,Farias}. So the presence of triose phosphate isomerase ensures a rapid isomerization kinetics of the monomer,
and the emergence and preservation of homochirality of carbohydrates must have been obtained in another way than the homochirality
in proteins \cite{Toxvaerd2018}.

To understand the emergence and preservation of homochirality  in   polysaccharides and  in the Glycolysis, one
must notice, that these consecutive chains of reactions are reversible.
Take for instance the central step in Glycolysis,
\begin{eqnarray}
	\textrm{D-fructose-1,6-diphosphate} \rightleftharpoons  \nonumber\\ 
	\textrm{dihydroxyacetone phosphate } \nonumber\\
	\textrm{ + D-glyceraldehyde-3-phosphate}.
\end{eqnarray}   

The standard  Gibbs free energy $\Delta G^{^{\ominus '}}$ = 23.97 kJ/mol for this reaction is strongly positive, but
 the reaction proceeds readily in the forward direction under intracellular conditions ($\approx 10-100 \mu$M) in a living cell. This is because
 the reaction Gibbs free energy
 is negative, $\Delta_r G<0$, due to the cleavage of fructose with an increased entropy! In contrast, the reverse reaction is spontaneous
at higher concentrations of
dihydroxyacetone phosphate and D-glyceraldehyde 3-phosphate
 with a gain of free energy by the synthesis of fructose and with a storage of free energy
in fructose-6-phosphate, and in the first component, glucose-6-phosphate, in glycolysis.

 The isomerization kinetics with the chiral carbon atom in glyceraldehyde
is no longer active after condensation of glyceraldehyde with dihydroxyacetone by which the original chirality is preserved during the condensation.
The pentoses and hexoses contain additional chiral carbon atoms,
created by this condensation and another synthesis with a keto-enol isomerization between them.
But the chirality at carbon atom no. 5 (no. 4 position for pentoses)
is not affected by these keto-enol isomerizations.

The chiral structures are not limited to simple polysaccharides either
in that the chains in RNA and DNA are all D-structures of phosphate esters of ribose and deoxyribose respectively.
D-ribose-5-phosphate has a metabolism associated with glycolysis, and together these facts may explain the origin and conservation
of a chiral carbohydrate world.

Carbohydrates in living organisms are catalyzed by stereo-specific enzymes in the Glycolysis.
This fact can also explain the emergence of homochirality in carbohydrates in a prebiotic environment.
If at the beginning of abiogenesis with the consecutive reactions in the $\it{total}$ metabolism,  was just one stereospecific polypeptide which only
lowers the activation energy for a reaction with the D structure but not for the L structure, then the metabolism with this enzyme will ensure
a homochiral D-carbohydrate world.  The triose phosphate isomerase will drain the pool of the central units, glyceraldehyde-3-phosphate, for the L-conformations and
 convert L-glyceraldehyde-3-phosphate in the pool
to D-glyceraldehyde-3-phosphate and
 ensure a local racemic equilibrium in the pool. And the stereospecific enzyme will use D-glyceraldehyde-3-phosphate in the synthesis. The net result will be a total dominance of D-carbohydrates.
 Such enzymes exist. The enzyme hexokinase (glucokinase),
 which catalyzes the first step in glycolysis, only catalyzes the phosphorylation of D-glucose,
 $\textit{but not L-glucose}$. The enzyme ribokinase, which
 phosphorylates D-ribose does not catalyze the corresponding phosphorylation of L-ribose \cite{Chuvikovsky}.

 The first step in the Glycolysis is the stereospecific phosphorylation of D-glycose to D-glycose-6-phosphate. 
RNA and DNA, are correspondingly phosphor polyesters, so the phosphorylation is central for the establishment of the metabolism as well as genetics.
An explanation of the role of phosphorylation in Abiogenesis is given by Westheimer \cite{Westheimer}, who argued that phosphorylation stabilizes
the biomolecules. The phosphorylation must have appeared at the establishment of a metabolism. If the stereospecific proteins 
hexokinase and ribokinase by changes were present 
at the phosphorylation of the carbohydrates,  it is sufficient to ensure the homochirality of carbohydrates in the metabolism.

$\textit{In summary:}$
The homochirality in carbohydrates in living organisms is obtained with the enzymes hexokinase and ribokinase with a stereospecific catalysis of the D-configuration. 
Catabolic enzymes remove L-carbohydrates. These enzymes are found in the
earliest living systems.
So the emergence of homochirality in peptides and carbohydrates is fundamentally different. The emergence of homochiral peptides is
obtained by chiral discrimination by the secondary $\alpha-$helix and $\beta-$sheet conformations, whereas
the metabolic consecutive chemical reactions with stereospecific enzymes, which catalyze D-carbohydrates, lead to the emergence of homochirality in carbohydrates.

 A prebiotic environment with peptides in their secondary conformation and with stereospecific catalysis of D-carbohydrates in the metabolism explains the emergence and stability of
 the homochiral carbohydrate world. But it also establishes an order in evolution at the start of abiogenesis. Creating a prebiotic peptide world
 with stereospecific enzymatic proteins is necessary for the stereospecific Glycolysis with a homochiral D-carbohydrate world. So
 homochirality in peptides 
appeared before the advent of homochirality in carbohydrates.

 \subsubsection{ B2. Emergence of the genetics}
Genetics presents the network of consecutive chemical reactions with the link between DNA and proteins.
The primordial evolution of life with the emergence of genetics has been
the subject of many scientific investigations \cite{Frenkel-Pinter2020,Bose2022} and theoretical modeling \cite{Abel2025}.
The complex molecular structures of even the simplest living organisms raise several questions about their formation and the order of evolution.
A recent proposal divides the primordial era into four
major eras: (i) prebiotic, (ii) FUCA, (iii) progenotes, and (iv) the age with cellular organisms \cite{Prosdocimi2023,Prosdocimi2024}.
Another major event in the evolution is the emergence of autopoietic organization, i.e., the emergence of  a system capable of producing and maintaining
itself by creating its own parts \cite{Igamberdiev2023}.
According to the present article, the emergence of homochirality appeared in the prebiotic age. The next age
 is FUCA (the first universal common ancestor), which is thought to be a non-cellular entity that was the earliest
 system with a genetic code capable of biological translation of RNA molecules into peptides to produce proteins.
The emergence of RNA nucleotides and catalytic RNAs is reviewed in \cite{Ruzov2025}. The emergence of DNA in the primordial era is described in \cite{Farias2023}.
However, all the cited articles about the emergence of genetics concern the complex molecular evolution of biomolecules and primordial organizations before 
the age with living organisms,
and none of the articles concern the emergence of homochirality, and  with good right, because this event with homochiral strings of D-Ribose units appeared in
the prebiotic era, prior to
FUCA according to the present hypothesis.

The emergence of homochirality in carbohydrates was explained in the previous section by chiral purification in the network of
chemical reactions in the metabolism,
and one can explain the emergence of genetics with
it's stereospecific chemical reactions using the same qualitative arguments.
But there is a relationship that can support this hypothesis, and it can be formulated as:\\
\textit{ Why a D-carbohydrate, L- amino acid world, and not a  L-carbohydrate, D- amino acid world?}

There is a funny fact with Pauling's Figure 1 for the $\alpha-$helix structure: it is a peptide with all D-units of amino acids!  \cite{Dunitz2001}.
Linus Pauling's anticlockwise, all D-peptide helices must be the key molecules in a  D-protein- and
L-carbohydrate world, a world which does not exist on planet Earth. As described in 2.1 and 2.2.1
it is, in fact, fully understandable that homochirality can be obtained from a racemic mixture of chiral molecules with isomerization kinetics.
 Chiral discrimination in favor of homochirality can lead to phase separation in a mixture
 with racemic composition \cite{Toxvaerd2000,Dressel2014},  and  the coarsening and growth of domains with
different chirality (L- and D-) will favor
the biggest domain and lead to a symmetry break with the dominance of one of the chiral conformations.
   \cite{Toxvaerd2000,Toxvaerd2009,Toxvaerd2017}.
However,  the compact proteins with $\alpha$-helix structures 
in an aqueous suspension are autonomous, and
the isomerization kinetics which, on the one hand, ensures homochirality in the proteins by the chiral discrimination
in the secondary structure,  will, on the other hand, give a statistically equal amount of $\alpha$-helixes 
with D- and L-units of amino acids, respectively. So this fact does not explain that we only have peptides with units of L-amino acid and  D-carbohydrides,  there
should also exist environments with the opposite chirality.

There is, however,  a crucial quality of the metabolism  and $\it{catabolism}$ of carbohydrates
and peptides, which   can explain  the emergence and preservation of 
 the L-protein and D-carbohydrate world only. 
The L-carbohydrates are not included in the Glycolysis  of carbohydrates, but simple bacterium  \cite{Shimizu} and procaryotes \cite{Yoshida} 
contains  enzymes, which catalyze a catabolic pathway for
L-glycose.
Bacteria contain enzymes with a catabolism of D-amino acids, D-amino acid dehydrogenase \cite{Olsiewski,Naganuma2018}.
The critical barrier to the growth and stability of one of the chiral worlds must be the establishment of its metabolism and catabolism:
\textit{ Who came first and dominates will ``eat'' the other by means of a 
catabolism!}

\textit{In conclusion}:
The emergence of metabolism and genetics evolved after the establishment of homochiral peptides in their secondary configurations,
Both prebiotic syntheses are a necessity for life and deserve  milestones before the
next step in the evolution of living organisms.

\subsection{C. The emergence of a cell with cell division}
The last milestone in the evolution of biomolecules and prebiotic environment(s) at the origin of life is the milestone for the emergence of a cell with cell division.
The cell division in a simple bacteria is a very complex process associated with molecular rearrangements and cell polarity before the division. Cell polarity
refers to the spatial differences in shape, structure, and function within a cell at the cell division  \cite{Dworkin,Buskila,Laloux,Brown,Orlando,Treuner2014,Kim}.
The polarity is associated with protein rearrangements within the cell. Most likely there must have been a prehistory to the present
 very complicated cell division in a bacteria and with a much simpler cell polarity.

As with the origin of homochirality, many articles have
concerned with the origin of cells with cell division.
Usually one explains the appearance of cells as non-equilibrium phase separation and coarsening, ''Ostwald Ripening'', between water and a
hydrophobic cell with biomolecules. One is again facing the problem of whether a cell with cell division could have a precursor and a
 ``template'' for phase separation and droplet formation before the emergence of a living cell with its cell division. 
A recent article describes such a template.
It is a simple  model with active chemical reactions which illustrates that the growth of a droplet 
with these reactions included in the phase separation and droplet formation can explain the division of
a growing droplet into smaller droplets \cite{Zwicker2017}. The kinetics for non-eqillibrium droplet
systems with chemical reactions is reviewed in \cite{Zwicker2022}. The experimental support for such an active domain division by chemical reactions of
biomolecules in droplets is reviewed in \cite{Soding2020}.

Bacteria with cell division in an aqueous solution is not a  simple two-phase binary system with a
hydrophobic phase of droplets with chemical reactions. The
interior of a bacteria is from a physicochemical point of view a cell with a diluted aqueous (cytosol) ionic suspension of highly ordered hydrophobic biomolecules and
with the chemical reactions and cell polarity in the cytosol. This feature article is, however, not focused on  the precise 
description of the emergence of cell with cell division, but on the order in which life evolves. It is difficult
 to imagine that the cell divisions at the emergence of cells took place before the metabolism and genetics were established. Rather it might be part of the
 kinetics which, in the model quoted above, ensured the cell division.

\textit{In conclusion}: There are many experimental examples in the reviews of
cell polarity and cell division  and in the articles cited above which state, that the kinetics in the cells with proteins  play a crucial and
 enzymatic role in cell polarity at the cell division. This fact indicates that the emergence of cells with cell division appeared
 after the emergence of a kinetics with homochirality in metabolism and genetics. The milestone at the appearance of cells with cell division
 must most likely be the last milestone in the prebiotic evolution.

\section{Summary: The locations of the milestones at the origin of life}
The origin of life is evolutionary with a network of spontaneous biochemical reactions, and the evolution has taken place over a very long time.
 The evolution of life contains, however some ``landmarks'' and bottlenecks, which in a revolutionary manner directed the evolution.
 No life like ours is without homochiral proteins and carbohydrate molecules, so the appearance of homochirality in metabolism and genetics
 are milestones in the evolution of life.
 There are, however, physicochemical and experimental facts that indicate that the emergence of homochirality in proteins and carbohydrates was
 fundamentally different. Homochirality in proteins was ensured by chiral discrimination in the secondary structure of the protein. In contrast, carbohydrate homochirality was obtained by the Formose synthesis of carbohydrates and stereospecific enzymes. If so, it indicates an order
 in evolution and places the first milestone at the origin of life  at the time when homochiral enzymes were present. And only then
 could the Glycolysis and the genetics be developed.
 In the same manner, the complexity of cell polarity for bacteria with enzyme-guided cell division indicates, that  a biochemical reaction-driven cell division
 with metabolism and genetics is the last milestone in the prebiotic evolution of life.

There are so many other remarkable structural and biochemical features of bacteria, which are not mentioned here  and which deserve
 a milestone upon their emergence,
 e.g. active channels in the membranes. But the milestones leave us with a problem: Where shall we place the stones? 

 The article \cite{Toxvaerd2023} places the location for the emergence of homochirality, metabolism, and genetics
 in the crust in the upper $\approx $ 100 km part of the lithosphere in the Earth's mantle. 
 Based on the latest literature regarding the conditions
in the Earth's crust and upper part of the mantle, several factors point to the crust being the
location for the prebiological self-assembly of biomolecules, and nothing is against it. 
 The crust and the upper part of the mantle  contain a substantial amount of water \cite{Ohtani2021},
  The crust also contains all the chemical components \cite{Wedepohl1995,Harrison2020}, necessary for the 
 synthesis of amino acids \cite{Seitz2024}, simple carbohydrates in the Formose reactions, the polymerizations of peptides and carbohydrides and
 the more complex biomolecules. 
 At the time before
the emergence  of life the crust  beneath hydrothermal locations   contained  an aqueous environment
   and it is a more likely place
for the prebiological synthesis over millions of years of peptides and carbohydrates, and the establishment of homochirality by polymerization.
 So the first three milestones for the emergence of homochirality and mortality, the metabolism, and genetics 
 shall be placed somewhere in the crust, but not the
 last milestone for the emergence of cells with cell division.

  Before the milestone for the emergence of cells with cell division is placed one should notice where on Earth the oldest examples of
 cellular life are found.  It is in ``hot spring'' environments \cite{Djokic2017,Dodd2017,Cavalazzi2021}.
 There are bacteria in the hot springs, as well as in the crust \cite{Li2020}.
 The deposits in hot springs are reviewed in \cite{Marais2019}. The theories and models for droplet formation and division are
 all for non-equilibrium quenched systems into phase points with phase separation. A thermodynamic quench expresses a rapid change
 of a system's phase point by changing the thermodynamic variables: temperature, pressure, and composition. Such a quench will appear if a prebiotic
 (open) systems in the crust with metabolism and genetics are exposed to a quench at the interface between the crust and the hot spring environment
 in the Hadean Ocean. The milestone for the emergence of cells with cell division should been placed there for $\approx$ 4 Gyr ago.

A wet crust is a suitable place for the biochemical evolution primordial stages. Here we have advocated for the Earth's crust  in the
Hadean Eon
as the most likely place,  but it might as well have been on Mars. The requirement for a moderate temperature for the synthesis of stable biomolecules
sets a limit to how soon after the emergence of the solar system such a biosynthesis could take place, and this fact is in favor of Mars, which is a
terrestrial planet similar to Earth. It was created at the same time as the Earth but cooled down before the Earth \cite{Plesa2018}.

\section{Verification}
The proposed sequence of some critical events in Abiogenesis, starting with the milestone at the emergence of homochirality in peptides,
is based on the hypothesis that peptides are stable in water with low water activity but unstable in the cytosol.
The many experimental observations of aging in biosystems, as well as the thermodynamics for the peptide synthesis, indicate this.
The hypothesis can be tested by direct experiments of the stability of the secondary configuration, and the
homochirality of peptides as a function of the water activity. However, if the instability of proteins in biosystems
appears over a lifetime (e.g., cataracts), it may be necessary  to increase the temperature in \textit{in vitro} experiments
to overcome the reaction energy barrier in the reversible chemical reactions related to the establishment and stability
of the secondary structures of proteins.

\section*{Funding}
This work was supported by the VILLUM Foundation’s Matter project, grant No. 16515.

\section*{Declaration of Interest Statement}

The author declare no compeeting interest.


\end{document}